\documentstyle[12pt]{article}
\begin{document}
\begin{center}
{\Large\bf SOME COMMENTS ON THE SPIN OF THE CHERN-SIMONS VORTICES }\\[2cm]
{\bf R. Banerjee}\footnote{rabin@boson.bose.res.in}\\
{\normalsize S.N.Bose National Centre for Basic Sciences}\\
{\normalsize Block JD, Sector III, Salt Lake City,
Calcutta 700091, India}\\
{\normalsize and}\\
{\bf P. Mukherjee}\\
{\normalsize A.B.N. Seal College}\\
{\normalsize Cooch Behar, West Bengal}\\
{\normalsize India}
\end{center}
\vspace{1cm}

\begin{abstract}
We compute the spin of both the topological and nontopological solitons
of the Chern - Simons - Higgs model by using our approach based on constrained
analysis. We also propose an
extension of our method to the non - relativistic Chern - Simons models.
The spin formula for both the relativistic and
nonrelativistic theories turn out to be structurally identical. This form
invariance manifests the topological origin of the
Chern - Simons term responsible for inducing fractional spin.
Also, some comparisons with the existing results are done.
\end{abstract}

\newpage
 In (2+1) dimensional field theories the gauge - field dynamics may be
dictated by the Chern - Simons (C-S) piece instead of the
 Maxwell ( Yang - Mills)
term. Interest in the C-S field theories originated from the observation
that the C-S term implements the Hopf interaction in (2+1) dimensional O(3)
nonlinear sigma model within the framework of local gauge theory \cite {wil1}.
 The
soliton sector of the theory offers excitations
(baby skyrmions) carrying
fractional spin and statistics \cite {wil2, bow}. Pure C-S coupled field
 theories can
support self dual vortex configurations - a fact exemplified by numerous
models \cite {dun}.
A remarkable aspect of the C-S gauge field is that it can be
coupled with both Poincare and Galileo symmetric models
\cite {jac1, ban1, chak1, ban2, duv}. The latter
possibility is very useful in view of the applications of the C-S theories
in condensed matter physics \cite {frad}. 
There are different
computations of the spin of the C - S vortices, both relativistic 
\cite{bow, hon, jac2, jac3, ban3, ban, ots} and non-relativistic \cite
{eza1,eza2}. However,
the results do not always agree \cite {not}.
Recently, we have proposed a general framework for obtaining the spin of 
relativistic C - S vortices \cite{ban3} which is found to give consistent
results for a variety of C - S theories \cite{ban}. In this paper an extension
of our method \cite{ban3,ban} applicable to the nonrelativistic C - S 
models is presented. Consequently, a general method, equally viable in the 
relativistic and non-relativistic cases, is provided.
 To put our work in
the proper perspective let us first give a brief digression.

      The usual method of defining the spin of the C-S vortices is to identify
the spin with the total angular momentum in the static soliton configuration
 \cite {dun, hon, jac2, jac3}.
This angular momentum integral is constructed from the symmetric
energy - momentum (E - M) tensor obtained by varying the action with respect
to a background metric. Since this E - M tensor is relevant in
formulating the Dirac -
 Schwinger conditions \cite {sch} it will henceforth
be referred as the Schwinger
E - M tensor. Correspondingly, the angular momentum following
from this energy momentum tensor usually goes by the name of
Schwinger. It is both symmetric and gauge invariant, and also
occurs naturally in the context of the general theory of relativity.
For these properties it is also interpreted as the physical angular momentum.
Contrary to the Noether angular momentum, however,
the Schwinger
angular momentum does not have a natural splitting into an orbital and a
spin part \cite {wie}. Thus it is not transparent how the value of this
angular momentum
in the static limit may be identified with the intrinsic spin of the vortices.

       In the alternative field - theoretic definition of the spin of the
C-S vortices advanced in
 \cite {ban3} 
,one abstracts the canonical part
from the physical
angular momentum.
The canonical part is obtained by using the conventional
Noether definition. Both the canonical as well as the physical angular
momentum are obtained from the improved versions of the corresponding
E - M tensors to properly account for the constraints of the theory.
Now the Noether angular momentum contains the
orbital part plus the contribution coming from the spin degrees of freedom
as appropriate for generating local transformations of the fields under
Lorentz transformations \cite {bog}.
Its difference from the physical angular momentum
 is shown to be a total boundary containing the C-S gauge field only,
the value of which depends on the asymptotic limit of the C-S field. It
vanishes for nonsingular configurations. However, for the C-S vortex
configurations we get a nonzero contribution. This contribution is found to be
  independent of the origin of the
 coordinate system. It is possible to interpret this difference as an internal
angular momentum characterising the 
intrinsic spin of the C-S
vortices. The connection of the anomalous spin with the topological
C-S interaction is thus clearly revealed.

	The C-S coupled O(3) nonlinear sigma model provides an ideal example
for the comparision of the different methods. The Schwinger's E - M tensor
for the model is given by the well known expression \cite {wil1},
\begin{equation}
   \theta^s_{\mu\nu} = {2 \over f^2}(2\partial_\mu n^a\partial_\nu n^a -
			g_{\mu\nu}\partial_\alpha n^a\partial^\alpha n^a),
\end{equation}
where $n^a ( a = 1, 2, 3 )$ are the sigma model fields satisfying
\begin{equation}
     n^an^a = 1
\end{equation}
The angular momentum integral is,
\begin{equation}
      J = \int d^2x \epsilon_{ij}x_i\theta_{0j}^s
\end{equation}
Since the (0,j) component of $\theta_{\mu\nu}^s$ explicitly involves a time
derivative of $n^a$, J vanishes in the static configuration.
 The definition of \cite {hon, jac2, jac3}
 then predicts
zero spin of the baby skyrmions. However,
it is definitely proved from quite general arguments that these excitations
carry fractional spin and statistics \cite {wil2, bow}.
In fact it has been shown that the spin value is given by,
\begin {equation}
     S = {\theta \over {2\pi}}
\end  {equation}
where $\theta$ is the C-S coupling strength. The connection of the baby
skyrmions with the quasiparticles found in the quantum Hall state has been 
established, where the anomalous spin of the excitations play a crucial
role \cite{nay}. So the vanishing spin of the baby skyrmions predicted 
by the definition of \cite{hon,jac2,jac3} is clearly a contradiction.
This contradiction,
 which was earlier noticed in \cite {ots}, prompted alternative definitions
\cite{ban3, ots} for obtaining the spin of the vortices. In this context
      it is interesting to note that the definition of \cite {ban3}
 gives an entirely
satisfactory result for this model.

      In \cite {ban3} we work in a gauge -
independent formalism \cite {ban4} and the Schwinger E - M tensor is
required to be
augmented by appropriate linear combinations of the constraints of the
theory \cite {dir}, so as to generate proper transformation of the fields
\cite {han}.
 The difference of the angular momenta obtained from the Schwinger and the
canonical E - M tensors is
\begin{equation}
      K = -\theta\int d^2x \partial_i(x^iA^jA_j - x^jA^iA_j),
\end{equation}
where $A^i$ is the C-S gauge field.
The asymptotic form of the gauge field is only required to
compute K and this form is dictated by the requirements of rotational
symmetry and the Gauss constraint of the theory to be,
\begin{equation}
      A^i(x) = {Q \over 2\pi}\epsilon_{ij}{x^j \over{{\vert x\vert}^2}},
\end{equation}
where Q is the topological charge. Using (5) we get
\begin{equation}
      K = {Q^2 \over{2\pi}}\theta.
\end{equation}
For the baby skyrmions Q = 1 and the spin value agrees with
equation (4).
 This example shows
that the definition of \cite {ban3}
yields results compatible with general arguments \cite {wil2}.

      We now extend our formalism for
calculating the spin of the vortices in nonrelativistic models with C-S
coupling, which is the main thrust of the paper. 
The Galileo invariant models cannot be made generally covariant
and so a gauge invariant E - M tensor cannot be constructed along
the lines of Schwinger.
Nevertheless we can build up a gauge invariant E - M tensor using the
equations of motion \cite {jac1}. We are then able to apply our formalism
developed for the relativistic theories by substituting the
Schwinger E - M tensor
with this one. The resulting spin formula comes out to be exactly equivalent
to that obtained in the relativistic case, revealing the topological
connection of the origin of the fractional spin. Before
discussing the nonrelativistic case, a quick survey of a relativistic
example is appropriate.

The Lagrangian of the C-S-H model is
\begin{equation}
{\cal L} = (D_{\mu}\phi)^{*} (D^{\mu}\phi) + \frac{k}{2}
\epsilon^{\mu\nu\lambda} A_{\mu}\partial_{\nu}A_{\lambda} -
V(\vert\phi\vert)
\end{equation}
where the covariant derivative is defined by,
\begin{equation}
D_{\mu} = \partial_{\mu} + ieA_{\mu}
\end{equation}
This model is known to possess both topological as well as
nontopological vortices.
According to our definition \cite {ban3} the spin of the vortices is defined as
\begin{equation}
K = J^S - J^{N}
\end{equation}
where $J^S$ and $J^N$ are, respectively, the Schwinger and Noether angular
momentum. We work in a gauge independent formalism where the
constraints of the theory are weakly zero. Different
E-M tensors are thus required to be augmented
by appropriate linear combination of the constraints of
the theory to obtain proper transformation of the fields.

 From a detailed analysis of the model (8) we arrive at the following
augmented expressions for $J^S$ and $J^N$ \cite {ban},
\begin{eqnarray}
J^S &=&\int d^2{\bf x} \epsilon^{ij}x_i[\pi\partial_j\phi+
\pi^*\partial_j\phi^* - k\epsilon^{lm}A_j\partial_lA_m]\\
J^N &=&\int d^2{\bf x}[ \epsilon^{ij}x_i(\pi\partial_j\phi+
\pi^*\partial_j\phi^* - { k\over 2}\epsilon^{lm}A_l\partial_jA_m) +
	       {k \over 2}A^jA_j]
\end{eqnarray}
where $\pi(\pi^*)$ is the momentum conjugate to $\phi(\phi^*)$.
Using (11) and (12) in (10) we get
\begin{equation}
K =  - \frac{k}{2}\int d^{2}\vec{x} \partial^{i}[x_{i}A_{j}A^{j}
- A_{i}x_{j}A^{j}]
\end{equation}
which is the same formula as obtained for the O(3) nonlinear sigma
model ( see equation (5) ).
Note that the integrand is a boundary term so that only the
asymptotic form of the gauge field $A_i$ is required for the
computation of K.

For topological vortices, the matter field $\phi$ at infinity
bears a representation of the broken U(1) symmetry,
\begin{equation}
\phi \approx ve^{in\theta}
\end{equation}
where n is the topological charge.
The requirement of finite energy of the
configuration dictates that asymptotically,
\begin{equation}
eA_i = n\epsilon_{ij}{{x^j}\over{\vert x\vert}^2}
\end{equation}
The above form is rotationally symmetric and satisfies the Gauss law.
As a consequence the magnetic flux $\Phi$ is quantised
\begin{equation}
     \Phi = {{2\pi n} \over{e}}
\end{equation}
 After a straightforward calculation using (13) and (15), we obtain,
\begin{equation}
K = {{\pi k n^2}\over{e^2}}
\end{equation}

The nontopological vortices lie at the threshold of stability \cite {jac3}.
For these the magnetic flux $\Phi$ is arbitrary. The asymptotic form
of the gauge field is now expressed as
\begin{equation}
      A_i = {\Phi \over {2\pi}}\epsilon_{ij}{x^j \over{{\vert x\vert}^2}   }
\end{equation}
and the spin computed from (13) is
\begin{equation}
K = {{ k \Phi^2} \over {4\pi}}
\end{equation}

    Equations (17) and (19) give the spin of the topological and nontopological
vortices of the C-S-H model respectively. Notice that the sign of the
spin is +ve in both the cases, which again is the same as that of the
elementary excitations
of the model. In earlier analysis \cite{jac3} there was a
difference in sign  which was explained by the
introduction of a  new interaction. This is
not necessary in the present discussion.

Now we will apply the same general method to nonrelativistic
models. Consider the Lagrangian
\begin{equation}
{\cal L} = i\phi^*D_t\phi - {1\over{2m}}(D_k\phi)^*(D_k\phi) +
	     {k\over 2}\epsilon_{\mu\nu\lambda}A_\mu\partial_\nu A_\lambda
\end{equation}
where $\phi$ is a bosonic Schrodinger field. The model (20) is invariant
under the Galilean transformations and not under the transformations of
the Poincare group. Note that the Galilean transformations take time and
space on an unequal footing. So space-time metric is not defined. In
writing (20) we adopt a spatial Eucledian metric, covariant and contravariant
components are thus not to be distinguished.

The action of the model (20) cannot be made generally covariant.
The powerful
method of constructing a gauge invariant energy - momentum (E - M) tensor,
as formulated by Schwinger,
is thus not available. Nonetheless, it is possible to
 construct a gauge invariant
E - M tensor by appealing to the equations of motion
\cite {jac1}. Our program is then
clear. We will find a gauge-invariant momentum density from the matter
current obtained by using the equations of motion. These equations
will then be exploited to show the conservation of the corresponding momentum.
We work in the gauge independent formalism in contrast with the gauge-fixed
approach of \cite {jac1}. A
suitable linear combination of the Gauss constraint is to
be added with the gauge invariant momentum operator, in order to
generate the correct transformation of the fields under spatial translation.
A gauge invariant angular momentum is then constructed using this momentum
density. The canonical angular momentum obtained by Noether's prescription
is now subtracted from it. The spin of the vortices is, as usual,  defined by
\begin{equation}
K=J-J^N
\end{equation}
which is exactly similar to equation (10) with
the exception that J is now the gauge invariant angular momentum constructed
by using the equations of motion.

 From the Lagrangian (20) we write the Euler-Lagrange equation corresponding
to $A_\mu$,
\begin{equation}
k\epsilon_{\mu\nu\lambda}\partial_\nu A_\lambda = j_\mu
\end{equation}
where $j_\mu$ is given by
\begin{eqnarray}
j_0& =& \phi^*\phi\\
j_i& =& {1 \over{2im}}[\phi^*(D_i\phi) - \phi(D_i\phi)^*]
\end{eqnarray}
Observe that (22) leads to a continuity equation
\begin{equation}
\partial_0 j_0 +\partial_i j_i = 0
\end{equation}
so that
$j_0$ and $j_i$ can be interpreted as the matter density and
current density respectively.

 From the E-L equation corresponding to $A_0$ we get the Gauss constraint
of the theory
\begin{equation}
G=\phi^*\phi - k\epsilon_{ij}\partial_i A_j \approx 0
\end{equation}

Now we come to the construction of the gauge invariant momentum operator.
The (0-i)th component of the EM tensor $T_{0i}$(i.e. the momentum density)
is obtained from the matter current
\begin{equation}
T_{0i} = {i \over 2}[\phi^*(D_i\phi) - \phi(D_i\phi)^*]
\end{equation}
We verify by a straightforward
 calculation using the equations of motion that
$T_{0i}$ indeed satisfies the appropriate continuity equation,
\begin{equation}
\partial_0 T_{0i} + \partial_k T_{ki} = 0
\end{equation}
where $T_{ki}$ is the stress - tensor,
\begin{equation}
T_{ki} = {1 \over {2m}}[(D_k\phi)^*(D_i\phi) + (D_k\phi)(D_i\phi)^*
	  - \partial_i (\phi^*D_k\phi + \phi (D_k\phi)^*)]
\end{equation}
Using the expression(27) of $T_{0i}$ we construct a gauge invariant
momentum operator
\begin{equation}
P_i = \int d^2{\bf x}T_{0i}
\end{equation}
Exploiting
 (28) and neglecting the boundary term we find that $P_i$ is
indeed conserved,
\begin{equation}
{dP_i \over dt} = 0
\end{equation}
The boundary term vanishes due to the condition that the
covariant derivative $D_i\phi$ is zero on the boundary which is
required to keep the energy
 finite. 

The lagrangian (20)
is first order in time derivatives. It is then easy to read off the 
basic brackets from (20) by symplectic arguments. The nontrivial 
brackets are
\begin{eqnarray}
\{\phi({\bf x}),\phi^{*}({\bf y})\} &=& - i\delta({\bf x} - {\bf y})\nonumber\\
\{A_{i}({\bf x}),A_{j}({\bf y})\} &=& {1\over k}\epsilon_{ij}
\delta({\bf x} - {\bf y})
\end{eqnarray}
Using these basic brackets we obtain,
\begin{equation}
\{\phi({\bf x}), P_i\} = \partial_i\phi({\bf x}) + iA_i\phi({\bf x})
\end{equation}
Hence the transformation of $\phi$ deviates from the expected canonical
structure.
For proper transformation of the fields under spatial
translation we require to supplement $T_{0i}$ by the Gauss constraint,
\begin{equation}
T_{0i}^T = T_{0i} + A_iG
\end{equation}
and the corresponding momentum operator
\begin{equation}
P_i = \int d^2{\bf x}[{i \over 2}(\phi^*D_i\phi - \phi(D_i\phi)^*)
	+ A_i G]
\end{equation}
turns out to be an appropriate generator of spatial translation.
The term containing the Gauss operator in (35) 
exactly generates a piece in $\{\phi({\bf x}),P_i\}
$ which cancels the anomalous term in (33).

We now come to the construction of $J$, the gauge invariant angular
momentum, from the momentum density (35),
\begin{equation}
J = \int d^2{\bf x}\epsilon_{ij}x_i[{i \over 2}(\phi^*D_j\phi -
	     \phi(D_j\phi)^*) + A_j G]
\end{equation}
The canonical angular momentum $J^N$ is obtained from Noether's theorem
as \cite {chak1},
\begin{equation}
J^N = \int d^2{\bf x}[\epsilon_{ij}x_i({i \over 2}(\phi^*\partial_j\phi -
	     \phi(\partial_j\phi)^*) -
	   {k \over 2}\epsilon_{mn}A_m\partial_j A_n) +
	 {k \over 2}A_jA_j]
\end{equation}
Substituting (36) and (37) in (21) we obtain,
\begin{equation}
K = -{k \over 2}\int d^2{\bf x}\partial_i[x_iA^2 - x_jA_jA_i]
\end{equation}
Observe that the master formula for the calculation of spin (38) is identical
with equation (13). The asymptotic form of $A_i$ following from
general considerations already elaborated leads to the same
structure as in (15). Inserting this in (38) exactly reproduces
(17) as the spin of the vortices.

We note in passing that self - dual soliton solutions can be obtained by
including a quartic self - interaction in (20)
\cite {jac1}, which are
interpreted
as the nonrelativistic limit of the nontopological vortices of the
relativistic Chern - Simons - Higgs model considered previously.
The spin of these solitons can be calculated by (38) using the asymptotic
form (18). The result comes out to be identical with (19).
This is expected, because the existence of the fractional spin
is connected to
the Chern - Simons piece which is a topological term.
The spin of the vortices of the model (20) with quartic self - interaction 
and an external magnetic field was calculated earlier \cite{eza1,eza2}.
Their method was in spirit akin to that of \cite{hon,jac2,jac3} but
they had to subtract the background contribution to get the exact spin.
The result of \cite{eza1,eza2} scales with the topological number as in our
case, but with opposite signature. The same comments apply to this comparison
as made earlier in connection with the C - S - H model.

To conclude, we found that the usual method
of defining the spin as the static limit of the physical angular
momentum yields contradictory results when applied
to compute the spin of the solitons of the Chern - Simons (C-S)
coupled O(3) nonlinear sigma model\cite {ots}. In this connection we have
observed that a consistent result is obtained when we apply our general
formalism for computing the spin in the C-S theories \cite {ban3},
 which exploits
the constraints of the theory. Here the canonical part of the physical angular
momentum is abstracted by subtracting the canonical (Noether) angular
momentum from the angular momentum obtained from Schwinger's E - M tensor
\cite {sch}. The difference  was found to be nonzero for
singular configurations. In particular for  C-S vortices this
difference was shown to be independent of the origin of the coordinate
system. Consequently we interpret it as the intrinsic spin of the vortices.
The formula for the spin comes out
to be model independent
and
contrary to other approaches where detailed field configurations are
necessary, only the asymptotic form of the
gauge field is required for its evaluation.

    The spin of the topological and nontopological vortices
of the C-S-H model was reviewed by the general formalism mentioned
above. The spin of both types of vortices of the model comes out
with the same sign.
We also
find that the sign of the spin of the topological vortices is the same as
that of the elementary excitations of the model
\cite {dun2}. This is a
satisfactory result because the spin-statistics connection is then
respected with the usual Aharonov - Bohm phase. Notably, in \cite {jac3}
an opposite sign was found so that a new interaction was required to 
properly account for this phase \cite {kim}.

    Our formalism is directly applicable to the relativistic theories
but the Chern - simons interaction enjoys the rare distinction of
being suitable to be coupled to both Poincare and Galileo symmetric
models [5 -9]. We were thus motivated to extend our formalism
to the nonrelativistic theories. Moreover, a systematic
discussion of the spin in such theories is nonexistent.
Although an explicit calculation exists \cite {eza1,eza2}, the connection
of the method with the corresponding ones used for the relativistic models
is not quite clear. Our extended formalism treated the nonrelativistic 
models within the same general framework used for the relativistic case.
 The resulting spin formula was
identical with that obtained for the relativistic theories. This points
to the topological origin of the C-S term, responsible for the induction
of the fractional spin, either in relativistic or nonrelativistic
models.

{\bf Acknowledgements}

 One of the authors (PM) thanks Professor
C.K.Majumdar, Director of the S.N.Bose National Centre for Basic Sciences,
for allowing him to use the institute facilities.


\begin{thebibliography}{99}
\bibitem{not} Significantly, the values reported in [11] and [13]
also differ
in sign among themselves
notwithstanding the fact that they use the same definition of spin.
\end{thebibliography}
\end{document}